\documentclass[3p,times,twocolumn]{elsarticle}

\usepackage{amssymb}

\usepackage[figuresright]{rotating}

\newcommand{\isotope}[2]{$^{#2}{\rm #1}$}

\begin{document}

\begin{frontmatter}


\tnotetext[label1]{}
\ead{josephf@mit.edu}


\title{Project 8: Using Radio-Frequency Techniques to Measure Neutrino Mass}


\author{J.~A.~Formaggio\\ For the Project 8 Collaboration}

\address{Laboratory for Nuclear Science, Massachusetts Institute of Technology, Cambridge, MA 02139}

\begin{abstract}
The shape of the beta decay energy distribution is sensitive to the mass of the electron neutrino. Attempts to measure the endpoint shape of tritium decay have so far seen no distortion from the zero-mass form. Here we show that a new type of electron energy spectroscopy could improve future measurements of this spectrum and therefore of the neutrino mass. We propose to detect the coherent cyclotron radiation emitted by an energetic electron in a magnetic field. For mildly relativistic electrons, like those in tritium decay, the relativistic shift of the cyclotron frequency allows us to extract the electron energy from the emitted radiation. As the technique inherently involves the measurement of a frequency in a non-destructive manner, it can, in principle, achieve a high degree of resolution and accuracy.  
\end{abstract}


\end{frontmatter}


\section{Motivation}
\label{sec:motivation}

Ever since Enrico Fermi's original proposal\cite{fermi_tentativo_1933}, it has been known that the neutrino mass has an effect on the kinematics of beta decay.   Measurements have always suggested that this mass was very small, with
successive generations of experiments giving upper limits\cite{weinheimer_high_1999}\cite{bib:Troitsk}, most
recently ${m_\nu}_\beta < 2.3$ eV.  The upcoming KATRIN experiment anticipates having a sensitivity of 0.20 eV at 90\% confidence level\cite{bib:KATRIN,bib:Thummler}.  If the neutrino mass is much below 0.20 eV, it is difficult to envision any classical spectrometer being able to access it.  Oscillation experiments, however, tell us with great confidence that the tritium beta decay neutrinos are an admixture of at least two mass states. Data indicate that the effective mass must satisfy ${m_\nu}_\beta > 0.005$ eV under the normal hierarchy or  ${m_\nu}_\beta > 0.05$ eV in the inverted hierarchy.  These bounds provide a strong motivation to find new, more sensitive ways to measure the tritium beta decay spectrum.

To make advances toward lower and lower masses, it is important to develop techniques that allow for extremely precise spectroscopy of low energy electrons.  Current electromagnetic techniques can achieve order $10^{-5}$ in precision, but are at the limit of their sensitivity.  Therefore, a new technique must be pursued in order to approach the inverted or even the hierarchical neutrino mass scale implied by current oscillation measurements.  The proposed technique makes use of the radiation emitted during cyclotron motion in order to extract the energy of the electron ejected in tritium beta decay.  As the technique inherently involves the measurement of a {\em frequency}, it can, in principle, achieve a high degree of resolution and accuracy.  The combination of these two features makes the technique attractive within the context of neutrino mass measurements.

\section{Neutrino Mass via Beta Decay}

The most sensitive direct searches for the electron neutrino mass
up to now are based on the investigation of the electron spectrum
of tritium $\beta$-decay. The electron energy spectrum of tritium $\beta$-decay for a neutrino with
component masses $m_1, m_2,$ and $m_3$ is given by

\begin{eqnarray*}
{dN \over dK_e} \propto F(Z,K_e) \cdot p_e \cdot (K_e+m_e) \cdot (E_0-K_e) \cdot \\
\sum_{i=1,3}|U_{ei}|^2\sqrt{(E_0-K_e)^2-m_i^2} \cdot \Theta(E_0 - K_e - m_i)
\end{eqnarray*}
 
\noindent where $K_e$ denotes the electron kinetic energy, $p_e$ is the electron
momentum, $m_e$ is the electron mass, $E_0$ corresponds to the total decay energy, $F(Z, K_e)$
is the Fermi function, taking into account the Coulomb interaction
of the outgoing electron in the final state, $Z$ is the atomic number of the final state nucleus, and $U_{ei}$ is the element from the PMNS mixing matrix. As both the matrix elements and $F(Z,K_e)$ are independent
of the neutrino mass, the dependence of the spectral shape on $m_i$ is
given solely by the phase space factor. In addition, the bound on
the neutrino mass from tritium $\beta$-decay is independent of whether
the electron neutrino is a Majorana or a Dirac particle.

There currently exists two techniques to measure the beta decay spectrum with sufficient sensitivity to extract neutrino masses at the eV or sub-eV scale.  Calorimetric techniques, such as those employed by the MARE collaboration\cite{bib:MARE}, measure the total energy deposited by the decay.  In such cryogenic measurements, the source and detector are one and the same, providing a more favorable scaling.  The use of extremely long-lived isotopes, however, makes realizing sufficient rates difficult.  Spectrometer techniques, such as those in use by the KATRIN experiment\cite{bib:KATRIN} and its predecessors, separate the source from the detector in order to measure the energy of the electron to very high precision.  KATRIN very efficiently removes the bulk of the low energy spectrum.  However, the separation of the electron and the source requires a large-scale apparatus.  KATRIN is currently at the limit of such spectroscopic techniques, with a projected mass sensitivity of 200 meV.

We propose a third approach to neutrino mass measurements, using radio-frequency techniques to measure the beta decay energy spectrum.  The technique has the potential of retaining the high resolution available in spectroscopic measurements while removing the necessity of extracting the electron from the source.  A description of the technique is given below.

\section{Description of Technique}
\label{sec:technique}

Imagine a charged particle, such as an electron created from the decay of tritium, traveling in a uniform magnetic field $B$.  In the absence of any electric fields, the particle will travel along the magnetic field lines undergoing simple cyclotron motion.  The characteristic frequency $\omega$ at which it precesses is given by

\begin{equation}
\label{eq:cyfreq}
\omega = \frac{e B}{\gamma m_e} = \frac{\omega_c}{\gamma} = \frac{\omega_c}{1+\frac{K_e}{m_e c^2}},
\end{equation}

\noindent where $(\omega_c)$ is the cyclotron frequency and $\gamma$ is the relativistic boost factor. The cyclotron frequency, therefore, is shifted according to the kinetic energy of the particle and, consequently, any measurement of this frequency stands as a measurement of the electron energy.  Measurements of the relativistic cyclotron frequency shift have been made, but only within the context of electron traps.  Electron energies as low as 16 meV have been measured~\cite{bib:Gabrielse}.  In our case, we are interested in free-streaming electrons with energies near the endpoint of tritium beta decay.  These electrons have a kinetic energy of 18.6 keV or, equivalently, a boost factor $\gamma \simeq 1.0365$.

Since the relativistic boost for the energies being considered is close to unity, the radiation emitted is relatively coherent.  Each electron emits microwaves at frequency $\omega$ and a total power which depends on the relativistic velocity $\beta$\cite{bib:Johner}:

\begin{equation}
P(\beta,\theta) = \frac{1}{4\pi\epsilon_0}\frac{2q^2\omega_c^2}{3c}\frac{\beta_\perp^2}{1-\beta^2}
\end{equation}

\noindent where $q$ is the electron charge, $\epsilon$ is the permittivity of free space, and $c$ is the speed of light. For a magnetic field strength of 1  Tesla, the emitted radiation has a baseline frequency of 28 GHz.  This frequency band is well within the range of most commercially available radio-frequency antennas and detectors. It is conceivable, therefore, to make use of radio-frequency detection techniques in order achieve precision spectroscopy of electrons.  The typical power emitted by these electrons is sufficiently high to enable single-electron detection.

\begin{figure}[ht]
\includegraphics[width=1.00\columnwidth]{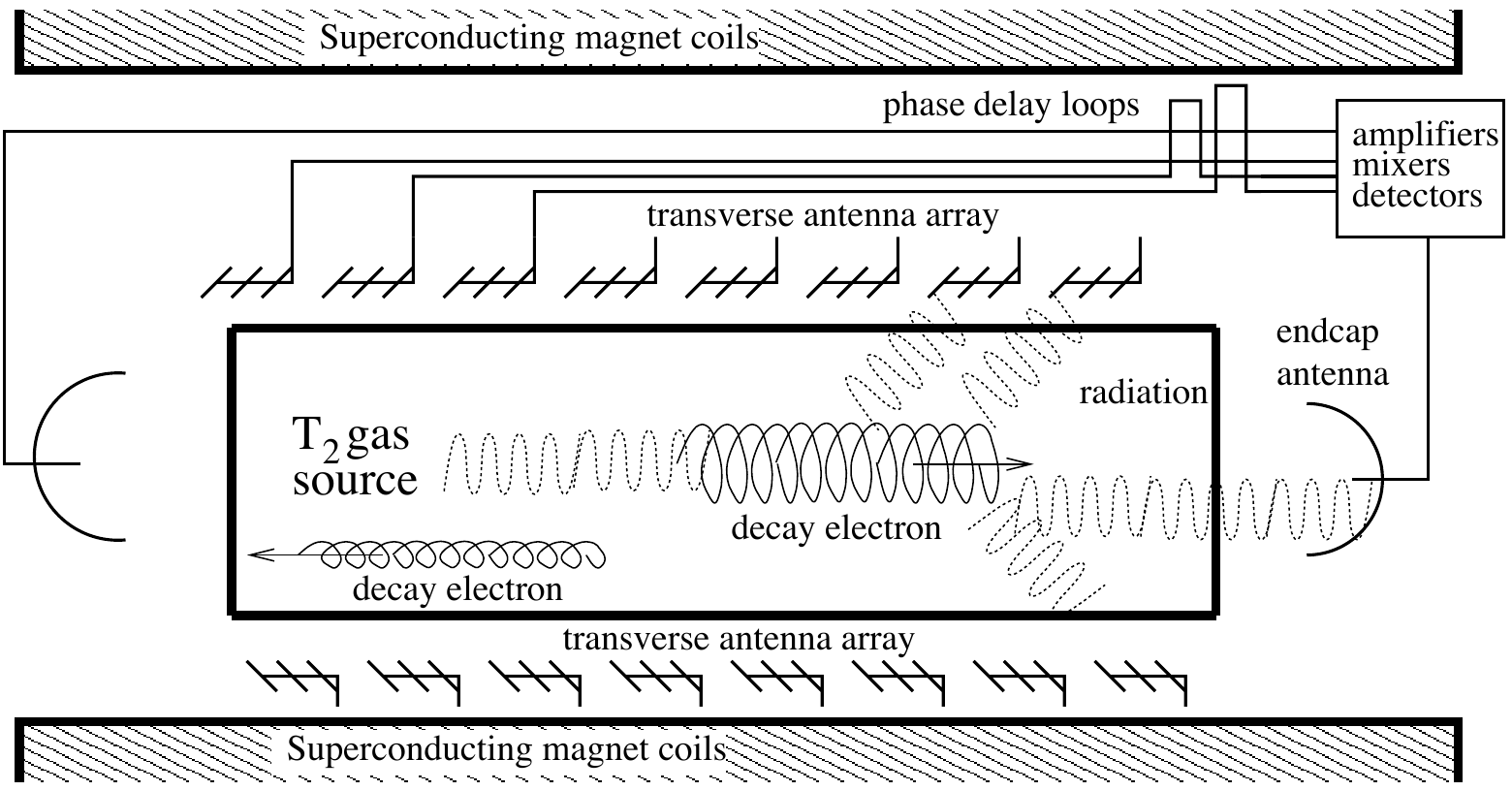} 
\caption{Schematic of a hypothetical tritium beta decay experiment.  A chamber encloses a diffuse gaseous tritium source under a uniform magnetic field.  Electrons produced from beta decay undergo cyclotron motion and emit cyclotron radiation, which is detected by an antenna array.}
\label{experiment}
\end{figure}

Consider the arrangement shown in Fig. \ref{experiment}.  A low-pressure supply of
tritium gas (electron source) is stored in a uniform magnetic field generated by a
solenoid magnet.  Tritium decay events release electrons with $0 < E_e < 18575$
eV (and velocity $0 < \beta < \beta_{\rm max}$ where $\beta_{\rm max} = 0.2625$) in random directions $\theta$ relative to the field vector.  The
electrons follow spiral paths with a velocity component 
($\beta_{||}$) parallel to the magnetic field.  Each electron emits microwaves at
frequency $\omega$ as defined in Eq.~\ref{eq:cyfreq} and a total power which depends on the perpendicular and parallel velocity components, $\beta_\parallel$ and $\beta_\perp$; respectively. By detecting the radiation and measuring its frequency spectrum, one obtains $\omega$ and hence the energy of the electron.

\begin{figure}[ht]
\includegraphics[width=1.00\columnwidth]{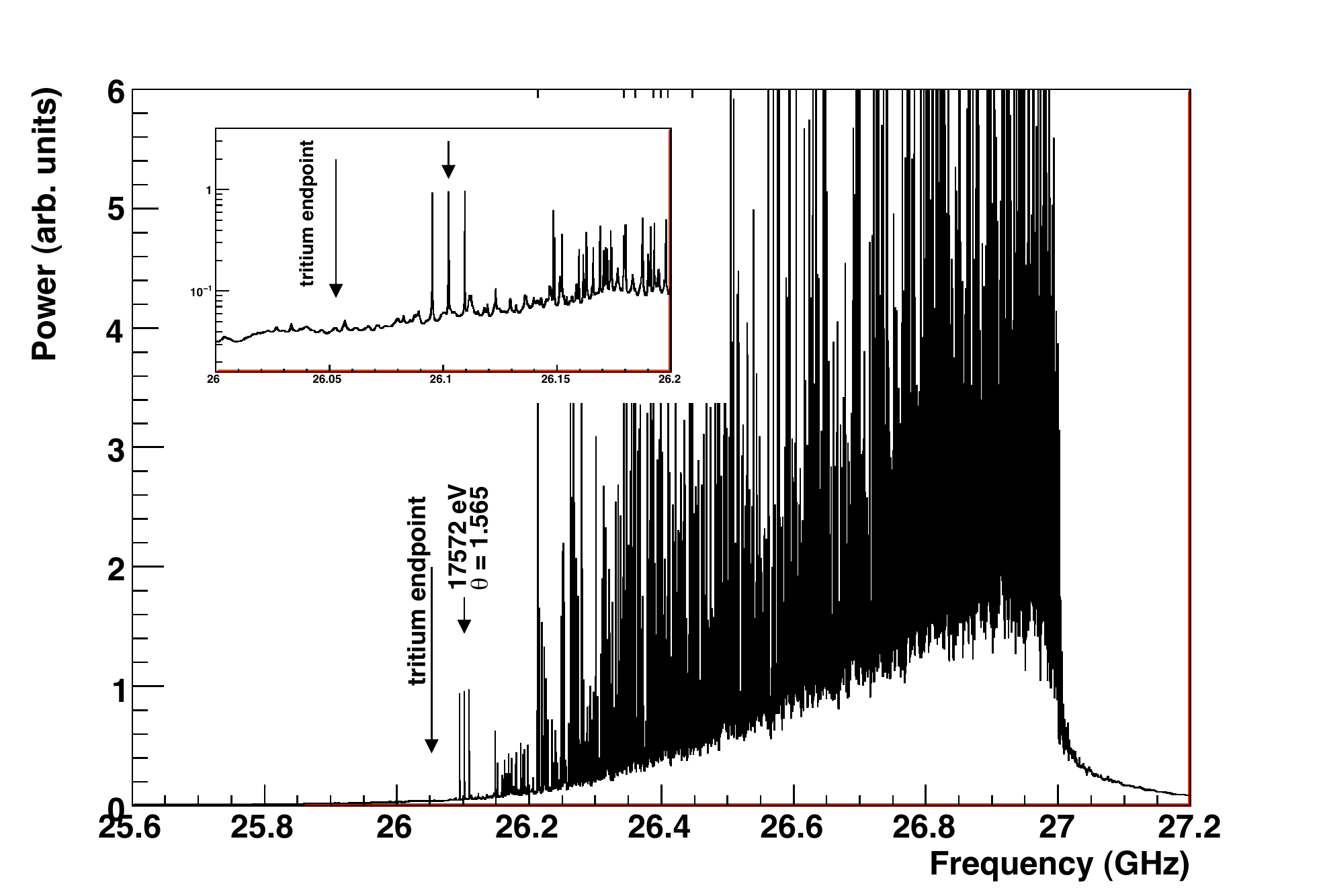}
\caption{Simulated microwave spectrum, showing the cyclotron emission of 10$^5$ tritium decays over 30$\mu$s in a 10m long uniform magnet ($\omega_0$ = 27 GHz, B $\sim$ 1 T) with a finely-spaced transverse antenna array.   e$^-$-T$_2$ scattering is neglected.  The short arrow points out a triplet of spectral peaks generated by an individual high-energy, high pitch angle electron.  The log-scale inset zooms in on this electron and the endpoint region.}
\label{spectrum}
\end{figure}

Although the emitted radiation is narrowband with frequency $\omega$, the signal seen in a stationary antenna is more complicated; generally it includes a Doppler shift due to $\beta_{\parallel}$, some dependence on the electron-antenna distance, and the differential angular power distribution of the emission.  The detected signal thus depends on the antenna configuration, and may have a nontrivial frequency content.   Nevertheless, one can consider a long array of evenly-spaced antennae oriented transverse to the magnetic field.  Any single transverse antenna may see the electron passing by, resulting in a complex, broadband ``siren'' signal which sweeps from blueshift to redshift.  However, the coherent sum signal from all of the antennae in the array must be quasi-periodic.  If the antennae signals are mixed appropriately, almost all of the complex Doppler effects sum incoherently while the unshifted cyclotron frequency remains coherent.  The final summed periodic signal appears as a ``carrier wave'' at frequency $\omega$ with (a) an amplitude modulation, because the antenna response varies periodically along the electron's path, and (b) possibly a small residual frequency modulation due to the relativistic ``beaming'' of the cyclotron radiation (see Figure~\ref{spectrum}(b)).  A more in-depth description of the technique can be found in Ref~\cite{bib:Monreal}.

In order to measure the electron energy to a precision $\Delta E$,
we need to measure the frequency to a relative precision of $\Delta f/f = \Delta E/m_e$.
For $\Delta E = 1$ eV this implies $\Delta f/f = 2\times 10^{-6}$.  In order to achieve a frequency precision of $\Delta f$, we need to monitor the signal for $t_{min} = 1/\Delta f$, according to Nyquist's theorem. This is a key number for several aspects of the experiment; for concreteness,
we discuss a reference design with a 1 T magnetic field and a $\Delta E=1.0$ eV energy resolution.  First, we want the beta electrons to have mean free flight times longer than $t_{min}$ (30 $\mu s$ in the reference design).   Due to T$_2$-e$^-$ scattering, this places a constraint on the density of the source.   The T$_2$-e$^-$ inelastic scattering cross section\cite{bib:aseev_energy_2000} at 18 keV is $\sigma = 3\times10^{-18}$ cm$^2$, so in order to achieve the appropriate mean free path the T$_2$ density cannot exceed $\rho_{max} = (t_{\rm min} \beta c \sigma)^{-1}$.   It also places a constraint on the physical size of the apparatus; we presume that our measurement ends when the particle reaches the end of some instrumented region.  If we want to be able to measure particles with minimum pitch angle $\theta_{min}$, the instrumented region needs to be of length $l = t_{min} \beta c \cdot \cos{(\theta_{\rm min})}$ long; in practice, engineering constraints on $l$ may set $\theta_{min}$.   Finally, $t_{min}$ also places a constraint on the magnetic field.  The electron continuously loses energy via cyclotron radiation; we want to complete our frequency measurement before it has lost energy $\Delta E$ due to radiative emission.   

One great advantage of the MAC-E filter technique used by experiments such as Mainz\cite{weinheimer_high_1999}, Troitsk, \cite{bib:Troitsk} and
KATRIN is the ability to reject effortlessly extremely
large fluxes of low-energy electrons, and to activate the detector and DAQ
only for the small fraction of decays near the endpoint.   A cyclotron
emission spectrometer will be exposed to all of the tritium decays in
its field of view (Fig. \ref{spectrum}); therefore, it is important that we be able to process these decays without unreasonable pileup.     

The main tool for separating signal from background is the
high-resolution and high-linearity nature of frequency domain
analysis.  Electrons with $E_e=0$ will generate fundamental signals at 27.992490 GHz; 18.575 keV electrons will emit fundamentals at about 27.010643 GHz; as each 1 eV analysis bin is about 50 kHz wide the full region-of-interest (ROI) is perhaps 1 MHz wide.  Detecting a narrow signal in the endpoint ROI is, by itself, insufficient to identify confidently an endpoint electron, since this band is also populated by the low-frequency sidebands of the much more numerous low-energy electrons; we will need to detect at least two spectral lines, possibly three, in order to confidently identify an electron.   In principle, any possible confusion source has a lower power than a real ROI source at the same frequency, but power measurements are expected to be fairly noisy in this system.  

Several other parameters conspire to reduce the impact of sideband confusion.  In order for a low-energy electron to put any sideband at all into the ROI, it must have a large $\beta_{\parallel}$ to generate the Doppler shift; however, a large $\beta_{\parallel}$ also leads to a quick exit from the spectrometer (and consequently a broad signal) and to lower emitted power.   Accidental coincidences may still occur.  If the detection criterion requires simply two high-power spectral peaks in coincidence, we estimate that a T$_2$ source strength of 10,000 Bq would give an accidental-trigger rate comparable to KATRIN's background event rate of one per $10^{13}$ effective source decays.  Requiring a third spectral peak raises this allowable source strength to approximately $10^{9}$ Bq.  

It is important that single electrons can be detected well above the noise level; first of all, to avoid false events from noise fluctuations; secondly, in order to approach as nearly as possible the Nyquist limit on the frequency resolution; thirdly, to increase the precision of total-power measurements and start/stop time estimates for each detected electron.  For our reference design with B=1 T, $\Delta E = 1 eV$, each resolution bin covers 50 kHz.  This bandwidth shows a thermal noise power of $6.5 \times 10^{-19}$ W/K, compared with a possible signal power in the neighborhood of $10^{-14}$ W.  In this frequency band there exist amplifiers which possess 10-20 K noise temperatures.  

A second noise source comes from the incoherent signals of non-endpoint and/or low-pitch beta electrons.  For our 1-T, 30 $\mu$s-analysis-period reference design, each 50 kHz analysis bin near 27 GHz will show approximately 10$^{-24}$ W/Bq of tritium noise.  This is compatible with robust signal detection in the presence of the $10^8$--$10^9$ Bq source allowed by pileup limitations.  We note, however, the possibility that this noise power will have non-Gaussian fluctuations.

This technique presents a very different systematic error budget than MAC-E filter experiments.  The spectrometer is continuously monitoring all decay energies, and thus is immune to slow source strength drifts.  We anticipate using an essentially static tritium gas whose electrostatic potential is fixed at both ends; this precludes large systematics due to source charging, voltage supply stability, flow-related Doppler shifts, and T$^-$ ion traps.  Microwave frequency measurements are easily stabilized against drifts at the $10^{-12}$ level.

\section{Next Steps}
\label{sec:Future}

\begin{figure}[ht]
\begin{turn}{-90}
\includegraphics[width=1.00\columnwidth]{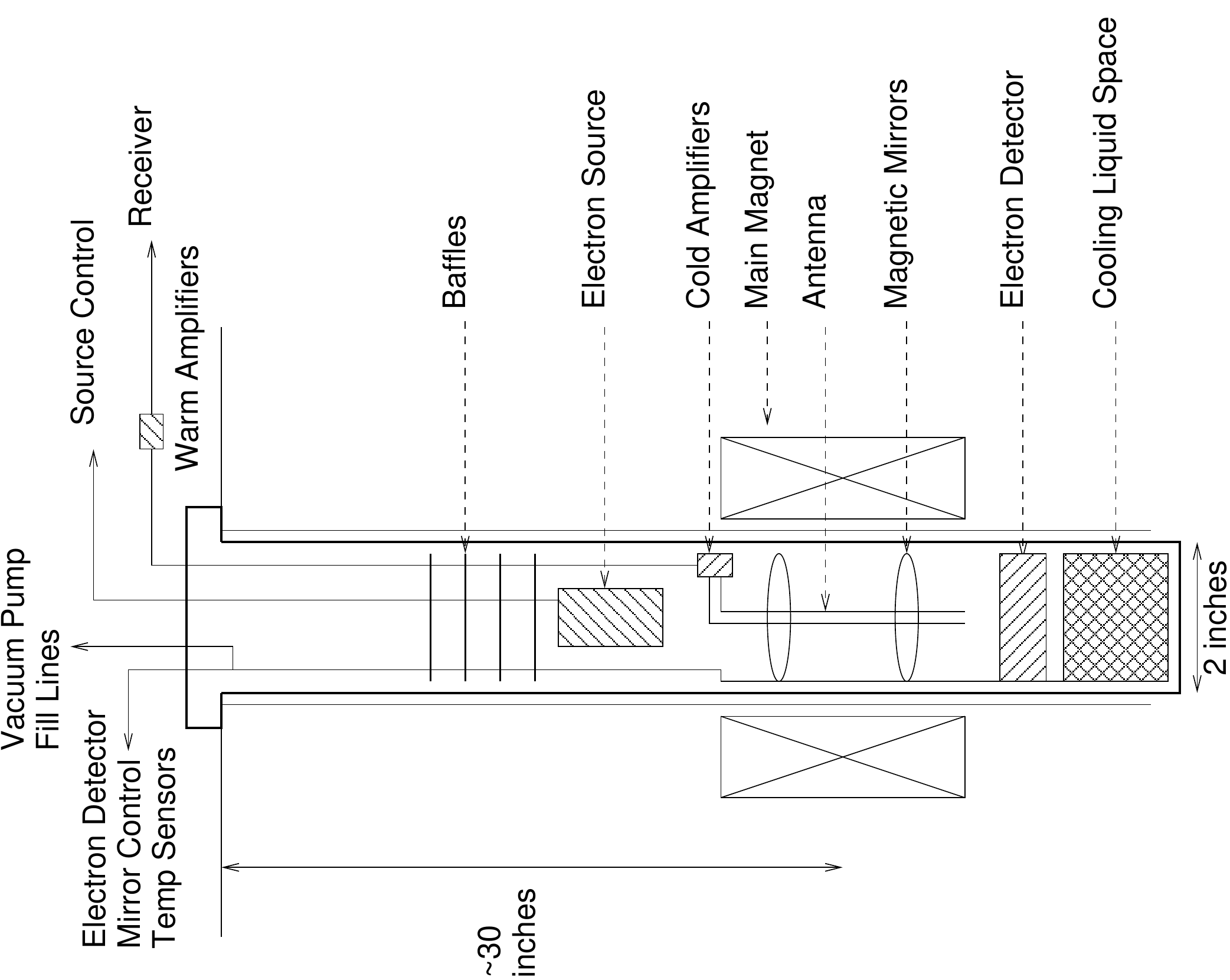}
\end{turn}
\caption{Schematic of the proposed research.  A large superconducting magnet encases a cold (4.4K) region where electrons from the source can be trapped and stored.  A series of wire antennas provide the coupling to the cyclotron frequency.  Signals are amplified and sent to the receiver electronics.  An electron detector is installed for monitoring purposes.}   
\label{fig:conceptual}
\end{figure}

We will initially operate a prototype system in a 1 T field, which corresponds to a baseline cyclotron frequency of 27.997 GHz.  A conceptual design of the prototype envisioned is shown in Figure~\ref{fig:conceptual}.  Each part serves one of two basic
functions, or both functions in some cases.  The first function is as a part
of the technology to be proven, i.e.,  the ability to detect the
cyclotron radiation of interest.  The second function is as part of a
system of calibrations and crosschecks to verify our understanding of
any signals observed, or to diagnose our inability to observe signals
in the case of null measurements.

The starting point is to have sources of electrons in the energy
region of interest which are well understood prior to their use in
this prototype.  The simplest source will generate single electrons in
a narrow range of energies and angles with respect to the magnetic
field.  The energy and angle should be independently controllable so
that the response can be mapped as a function of both variables.  Work
to create an electron gun with these capabilities is ongoing.  In
principle, the information gained from the simplest source is
sufficient to compute the response to any more complicated source.
The isotope \isotope{Kr}{83m} is an isotropically emitting source of
monoenergetic conversion electrons with $E = 17.8$\,keV.  The
observation and detailed quantitative understanding of the response to
a gaseous \isotope{Kr}{83m} serves as an excellent template to extrapolate to the more complex spectrum expected from tritium beta decay.

Electrons will need to be trapped in order to allow sufficient time
for a precise determination of their radiation frequency.  This will
be accomplished with a magnetic bottle: the trapping region will
be flanked by regions of increased magnetic field on both ends.  The
ratio of the maximum to the minimum field strength determines the range
of pitch angles confined in the trap.  Ability to switch the trap on
and off by ``opening'' either end will be necessary in order to limit the number
of electrons inside and limit ionization effects.

The total magnetic trap volume for the prototype will be approximately 1 mm$^3$ and trap electrons electrons with near maximal pitch angles ($\theta \ge 89^o$).  Though this is a considerable reduction in the allowed phase space, it optimizes the signal-to-noise ratio to facilitate electron detection.  A small two-wire antenna populates the trap volume for RF detection.  Signals are carried to a low noise cryogenic amplifier that can operate in the K$_a$-band.  The amplified signals are then transported outside the main volume and mixed to lower frequency for ready detection.  In addition to the observation of their cyclotron radiation, we also require a conventional method of detecting electrons, so as to monitor the \isotope{Kr}{83m} activity present in the test cavity.  This will allow the verification of the functionality of electron sources and the effectiveness of the magnetic trap.  Energy and time resolution are not critical requirements for this detector since its main purpose is to tag electrons and verify whether or not they were in the trap at a particular time.  It may be useful in early stages to use this detector as a trigger to initiate the search for a signal of cyclotron radiation.   However, another primary goal of this prototype is to demonstrate that the signal of cyclotron radiation is recognizable without any additional detection methods. A microwave source inside the prototype is also envisioned to serve as a power-frequency calibration.

\section{Summary}

Radio-frequency techniques are new in the field of neutrino mass measurements, but their application may provide a path to access sensitivity levels complementary to those already available from calorimetric and spectroscopic techniques.  An R\&D program is underway to determine the feasibility and sensitivity range of this technique.
\\





\bibliographystyle{elsarticle-num}
\bibliography{}



\end{document}